\documentclass[]{gPHT2e}

\begin{document}



\title{The absence of phase transition for the classical XY-model on Sierpi\' nski pyramid with fractal 
dimension $D$=2} 

\author{Michelle Przedborski and Bo\v zidar Mitrovi\' c$^{\ast}$\thanks{$^\ast$Corresponding author. 
Email: bmitrovic@brocku.ca\vspace{6pt}}\\\vspace{6pt}  {\em{Department of Physics, Brock University, 
St.~Catharines, Ontario, Canada L2S 3A1}}\\\vspace{6pt}\received{} }

\maketitle

\begin{abstract}
For the spin models with continuous symmetry on regular lattices  
and finite range of interactions the lower critical dimension is $d$=2. 
In two dimensions the classical XY-model displays Berezinskii-Kosterlitz-Thouless   
transition associated with unbinding of topological defects (vortices and antivortices).  
We perform a Monte Carlo study of the classical XY-model on Sierpi\' nski pyramids whose fractal 
dimension is $D=\log$4/$\log$2=2 and the average coordination number per site is $\approx$ 7. The specific
heat does not depend on the system size which indicates the absence of long range order. From the  
dependence of the helicity modulus on the cluster size and on boundary conditions we draw a conclusion 
that in the thermodynamic limit there is no Berezinskii-Kosterlitz-Thouless transition at any finite 
temperature. This conclusion is also supported by our results for linear magnetic susceptibility.  
The lack of finite temperature phase transition is presumably caused by the finite order of ramification 
of Sierpi\' nski pyramid.\bigskip

\bigskip

\end{abstract}

\section{Introduction}

One of the powerful predictions of the renormalization group theory of critical phenomena is universality 
according to which the critical behavior of a system is determined by: (1) symmetry group of the Hamiltonian, 
(2) spatial dimensionality $d$ and (3) whether or not the interactions are short-ranged \cite{gold92}.
The possibility of phase transitions on systems with nonintegral dimensionality $d$ was first considered by 
Dhar \cite{dhar} for the classical XY-model and for the Fortuin-Kasteleyn cluster model on the truncated 
tetrahedron lattice with the effective dimensionality 2$\log$3/$\log$5=1.365. No phase transition at any 
finite temperature was obtained. Subsequently in a series of papers Gefen {\em et al.} 
\cite{gab80,gab83,gasb84,gab84} examined the critical properties of the Ising model (discrete $Z_{2}$ symmetry) 
on fractal structures, which are scale invariant but not translationally invariant, in order to 
elucidate the relative importance of multiple topological factors affecting critical phenomena. 
They found that the Ising systems with given fractal dimension $D$ have transition temperature 
$T_{c}=$0 if the minimum order of ramification $R_{min}$, which is the minimum number of bonds that needs to be 
cut in order to isolate an {\em arbitrarily large} bounded cluster of sites, is finite. In the case of 
fractals with $R_{min}=\infty$  they presented arguments that $T_{c}>$0 for the Ising model. 
Monceau and Hsiao \cite{mh04} further studied the Ising model on fractals with $R_{min}=\infty$ using 
the Monte Carlo method and found weak universality in that the critical exponents depended also on 
topological features of fractal structures. 

For the models with continuous $O(n)$ symmetry, $n\geq$2, on fractal structures Gefen {\em et al.} 
\cite{gasb84,gab84} used a correspondence between the low-temperature properties of such models  
and pure resistor network connecting the sites of the fractal to argue that there is no long-range order at
any finite temperature if the fractal dimension $D<$2 even in the case of infinite order of ramification. 
Subsequently Vallat {\em et al.} \cite{vkb91} used the harmonic approximation to the XY-model ($O(2)$  
symmetry) on two-dimensional Sierpi\' nski gasket ($R_{min}$=3, $D=\log$3/$\log$2=1.585) to show that 
the energy of a vortex excitation is always finite and hence there is no 
Berezinskii-Kosterlitz-Thouless (BKT) transition \cite{gold92} at any finite temperature as  
free vortices are always present. These conclusions were confirmed in a recent Monte Carlo study of the 
full XY-model on two-dimensional Sierpi\' nski gasket \cite{mb10}.  

Here we present a Monte Carlo study of the XY-model on three-dimensional Sierpi\' nski pyramids 
with $R_{min}$=4 and fractal dimension $D=\log$4/$\log$2=2. The model is described by the Hamiltonian
\begin{equation}
\label{ham}
 H=-J\sum_{\langle i,j\rangle}\cos(\theta_{i}-\theta_{j})\>,
\end{equation}
where 0$\leq \theta_{i}<$2$\pi$ is the angle/phase variable on site $i$, $\langle i,j\rangle$ 
denotes the nearest neighbors and $J>$0 is the coupling constant. In the case of translationally 
invariant system in two dimensions this Hamiltonian gives rise to BKT transition and we investigate if the 
same holds in the case of a fractal structure with fractal dimension $D=$2.

The rest of the paper is organized as follows. In Section 2 we present our algorithm for generating 
Sierpi\' nski pyramids and outline the Monte Carlo procedure for 
calculating the thermal averages. Section 3 contains our results and discussion and in Section 4 we 
give a summary.  

\section{Numerical procedure}

The procedure which we used to generate three-dimensional Sierpi\' nski pyramids (SP) is illustrated in   
Figure 1, which shows the transition from the zeroth-order SP (the tetrahedron of unit side) to the
first order pyramid via translations by three nonorthogonal basis vectors ${\bf e}_{1}$=(1,0,0), 
${\bf e}_{2}$=(0,1,0) and ${\bf e}_{3}$=(0,0,1). The pyramid of order $m$+1 is obtained from the pyramid of 
order $m$ via translations by vectors 2$^{m}{\bf e}_{1}$, 2$^{m}{\bf e}_{2}$ and 2$^{m}{\bf e}_{3}$ ($m$=0,1,
$\dots$). It is clear that the number of vertices $N_{m}$ of the $m$th order Sierpi\' nski pyramid can be 
obtained from the recursion relation $N_{m}$=4$N_{m-1}$-6 with $N_{0}$=4. Thus, in generating the pyramid 
of order $m$+1 from the pyramid of order $m$ not every point of the $m$th order pyramid gets translated by 
{\em all} three translation vectors 2$^{m}{\bf e}_{1}$, 2$^{m}{\bf e}_{2}$ and 2$^{m}{\bf e}_{3}$. 
The top of the $m$th order pyramid, (0,0,0), is never translated. The remaining $N_{m}$-1 points are then 
all translated by 2$^{m}{\bf e}_{1}$. Next, the same points except for (2$^{m}$,0,0) are translated by 
2$^{m}{\bf e}_{2}$ and finally, all points except for (2$^{m}$,0,0) and (0,2$^{m}$,0) are translated by 
2$^{m}{\bf e}_{3}$. For pyramids of order $m\leq$9 we found it most efficient to represent a vertex $(i,j,k)$  
by the number $P=i$10$^{6}$+$j$10$^{3}$+$k$10$^{0}$ and the three translation vectors 2$^{m}{\bf e}_{1}$,  
2$^{m}{\bf e}_{2}$ and 2$^{m}{\bf e}_{3}$ by the numbers $T_{1}(m)=$2$^{m}\times$10$^{6}$, 
$T_{2}(m)=$2$^{m}\times$10$^{3}$ and $T_{3}(m)=$2$^{m}\times$10$^{0}$, respectively. Then the result of 
translating a point represented by number P by a vector represented by $T_{i}(m)$ is described by 
the number $P+T_{i}(m)$. From a given number $P$ representing a vertex it is easy to get its coordinates 
$(i,j,k)$ in the basis $\{{\bf e}_{1},{\bf e}_{2},{\bf e}_{3}\}$: $i=[P$/10$^{6}]$, where $[\cdots]$ 
denotes the integer part, $j=[(P-i\times$10$^{6})/$10$^{3}]$ and $k=P-i\times$10$^{6}-j\times$10$^{3}$. 

In the Metropolis Monte Carlo scheme of calculating the statistical averages for a model with only the 
nearest neighbor interactions it is necessary to provide the list of nearest neighbors for each site/vertex.
We take that all sites that are at a unit distance (the size of the edge of the elementary tetrahedron)
from a given vertex are its nearest neighbors. Thus the number of nearest neighbors of site $(i,j,k)$ varies 
with the order $m$ of Sierpi\' nski pyramid. For example, the vertex (1,1,0) has six nearest neighbors in 
the first order pyramid (Figure 1) but the same site has eight nearest neighbors in all higher order pyramids,
the additional two neighbors being vertices (2,1,0) and (1,2,0). For the Sierpi\' nski pyramids of orders 
$m=$4,5,6 we found the average number of neighbors per site to be 6.923, 6.981, 6.995 with standard 
deviations 0.903, 0.876, 0.869, respectively. Thus, the average coordination number for three-dimensional 
Sierpi\' nski pyramid is greater than the coordination number for three-dimensional simple cubic lattice.
This fact should be kept in mind when we discuss our numerical results in the next section. In constructing 
the list of nearest neighbors for the sites in a pyramid we found it convenient to group all sites 
according to the values of the sum $i+j+k$ of their coordinates (i,j,k). The sites with the same $i+j+k$
belong to the same plane parallel to the basal plane  of the pyramid defined by points (2$^{m}$,0,0), 
(0,2$^{m}$,0), (0,0,2$^{m}$), where $m$ is the order of the pyramid (see Figure 1). 
The nearest neighbors of a given site are located in the plane to which it belongs and in the 
neighboring planes. Throughout  
this work we employed two types of boundary conditions: closed, where the four corners of an $m$th order 
pyramid were considered to be coupled to each other and open, where the four corners are uncoupled to each 
other. 

The Monte Carlo (MC) simulation of the classical XY-model on Sierpi\' nski pyramids was based on 
Metropolis algorithm \cite{metro53}. We considered the pyramids of orders $m=$4 (130 sites), $m=$5 
(514 sites) and $m=$6 (2050 sites). For a pyramid of given order the simulation would start at a low 
temperature with all phases aligned. The first 120,000 steps per site (sps) were discarded, followed by 
seven MC links of 120,000 MC sps each. At each temperature the range over which each angle $\theta_{i}$ 
was allowed to vary \cite{bh88} was adjusted to ensure an MC acceptance rate of about 50\%. The errors
were calculated by breaking up each link into six blocks of 20,000  sps, then calculating the average 
values for each of 42 blocks and finally taking the standard deviation $\sigma$ of these 42 average 
values as an estimate of the error. The final configuration of the angles at a given temperature was used 
as a starting configuration for the next higher temperature.

\section{Numerical Results and Discussion}

The heat capacity per site shown in Figure 2 was calculated from the fluctuation theorem
\begin{equation}
\label{cv}
C=\frac{1}{N}\frac{\langle H^{2}\rangle - \langle H\rangle^{2}}{k_{B}T^{2}}\>,
\end{equation}
where $k_{B}$ is the Boltzmann constant, $T$ is the absolute temperature and $\langle\cdots\rangle$ 
denotes the MC average. The results did not depend on the type of boundary condition (closed or open) within 
the error bars in analogy to what was found for two-dimensional Sierpi\' nski gasket \cite{mb10}. In the 
same Figure we also show the results for heat capacity obtained for the XY-model on three cubic lattices  
with the periodic boundary conditions. The sizes of cubic lattices were chosen so that they are 
comparable to the sizes of three Sierpi\' nski pyramids. The peak in the specific heat of cubic 
lattices near $k_{B}T/J$=2.2 increases with system size \cite{lt89} as a consequence of diverging correlation  
length at a continuous (i.e. second order) phase transition. On the other hand the specific heat for the 
Sierpi\' nski pyramids is virtually size-independent indicating the absence of long range order for 
the XY-model on three-dimensional Sierpi\' nski pyramid at any finite temperature. Thus, although the 
average coordination number for the Sierpi\' nski pyramid ($\approx$ 7) is higher than its value for the 
cubic lattice, the topological properties of this fractal structure, in particular a finite order of 
ramification \cite{gasb84}, are responsible for the lack of long range order at finite $T$. The regular 
lattices have an infinite order of ramification as the number of bonds one needs to break in 
order to isolate an {\em arbitrary large} bounded cluster of lattice sites is infinite. For fractals with 
a finite order of ramification an arbitrarily large bounded cluster can be cut off from the rest of the 
structure by breaking off only a finite number of bonds and at finite temperature thermal fluctuations 
are sufficient to destroy the long range order.

The absence of
size dependence of the peak in $C$ is found for the XY-model in two dimensions \cite{tc79,VanHC81}. 
In that case 
the peak results from unbinding of vortex clusters \cite{tc79} with increasing temperature above    
the Berezinskii-Kosterlitz-Thouless (BKT) transition temperature $T_{c}$ where the heat capacity 
has an unobservable essential singularity \cite{bn79}. However the results in Figure 2 for fractal structures   
of fractal dimension $D=$2 do not necessarily imply BKT transition since size-independent peaks in $C$ could 
result from the average energy per site $\langle E\rangle$ changing monotonically from the values near 
-3.5$J$ (each site has about 7 neighbors on average) at low temperatures to near zero in 
high temperature paramagnetic phase. 

The main signature of BKT transition is a universal jump in helicity modulus $\gamma(T_{c})/T_{c}$=2/$\pi$ 
\cite{nk77} at critical temperature $T_{c}$. The helicity modulus $\gamma(T)$ measures the stiffness of the  
angles $\{\theta_{i}\}$ with respect to a twist at the boundary of the system. At zero temperature, when 
the angles are all aligned, one expects a finite value of $\gamma$ and at high temperatures, when the system 
is in disordered paramagnetic phase, $\gamma$ vanishes. For the XY-model on three-dimensional regular 
lattices $\gamma(T)$ decreases continuously with increasing temperature
and just below the transition temperature $T_{c}$ it obeys
a power law $\gamma(T)\propto |T-T_{c}|^{v}$ \cite{lt89}. In two dimensions, however, there is a 
discontinuity in $\gamma$ stemming from unbinding of the vortex-antivortex pairs at BKT transition from 
quasi-long-range order (order parameter correlation function decays algebraically) to disordered phase 
(order parameter correlation function decays exponentially).
In numerical simulations on finite systems discontinuity is replaced by continuous drop in $\gamma(T)$  
which becomes steeper with increasing system size (see, for example, \cite{mb10}). We should point out 
that since Sierpi\' nski pyramid is three-dimensional object topological defects are not only vortices 
in planes parallel to the faces of the pyramid (which are two-dimensional Sierpi\' nski gaskets) but also 
vortex strings. Kohring {\em et al.} \cite{ksw86} presented Monte Carlo evidence that a continuous 
phase transition for the XY-model in three dimensions is related to unbinding of vortex strings. 

We computed the helicity modulus for Sierpi\' nski pyramids following the procedure of Ebner and Stroud 
\cite{es83}. The idea is to think of Hamiltonian (\ref{ham}) as describing a set Josephson coupled 
superconducting grains in zero magnetic field, where $\theta_{i}$ is the phase of the superconducting 
order parameter on grain $i$. Then if a {\em uniform} vector potential ${\bf A}$ is applied 
the phase difference $\theta_{i}-\theta_{j}$ in (\ref{ham}) is shifted by 
$2\pi{\bf A}\cdot({\bf r}_{j}-{\bf r}_{i})/\Phi_{0}$, where ${\bf r}_{i}$ is the position vector of 
site $i$ and $\Phi_{0}=hc/2e$ is the flux quantum. The helicity modulus is obtained from the 
Helmholtz free energy $F$ as $\gamma=(\partial^{2}F/\partial A^{2})_{A=0}$, i.e.~
\begin{equation}
\label{hel}
\gamma=
\Big\langle\left(\frac{\partial^{2}H}{\partial A^{2}}\right)_{A=0}\Big\rangle-\frac{1}{k_{B}T}
\Big\langle\left(\frac{\partial H}{\partial A}\right)_{A=0}^{2}\Big\rangle+\frac{1}{k_{B}T}
\Big\langle\left(\frac{\partial H}{\partial A}\right)_{A=0}\Big\rangle^{2}\>.
\end{equation}  
We took ${\bf A}$ to be along one of the edges of elementary 
tetrahedron (Figure 1). Our results for $\gamma(T)$ obtained with closed boundary condition are shown in 
Figure 3. They are completely analogous to what was obtained for two-dimensional Sierpi\' nski gaskets 
\cite{mb10}:
a rapid downturn in $\gamma(T)$ starts near the universal 2/$\pi$-line but the low temperature values 
of $\gamma(T)$ decrease with increasing system size and the onset of the downturn, which is in the 
vicinity of the putative phase transition, shifts to the lower temperatures. For the XY-model on the 
square lattices, where BKT transition does occur, the low-$T$ values of helicity modulus and the onset of 
its downturn do not depend on the system size (see, for example, \cite{mb10}). Our results in Figure 3 
suggest that in thermodynamic limit $N\rightarrow\infty$ $\gamma(T)$ vanishes at any temperature $T>$0 
implying no BKT transition for the XY-model on Sierpi\' nski pyramid of fractal dimension $D=$2. This 
conclusion is reinforced by our results for $\gamma(T)$ obtained with the open boundary condition when 
four corners of an $m$th order pyramid are not coupled to each other, Figure 4. The helicity modulus is zero 
within the error bars which are larger than those obtained with the closed boundary condition. The results 
in Figures 3 and 4 indicate that closed boundary condition introduces additional correlations as was the 
case for two-dimensional Sierpi\' nski gaskets \cite{mb10}. 

Our conclusion about the lack of finite temperature BKT transition is supported by results for 
linear susceptibility
\begin{equation}
\label{susc}
\chi=\frac{\langle M^{2}\rangle-\langle M\rangle^{2}}{N^{2}k_{B}T}\>,
\end{equation}
where $M$ is the magnetization of the system, shown in Figure 5. For finite cubic lattices one gets a 
peak in $\chi$  near $k_{B}T/J=$2.2 whose size and sharpness increase with the number of sites as a result 
of diverging correlation length at the onset of long range order. In the case of Sierpi\' nski pyramids 
the peak in $\chi$ also grows with increasing system size but it also shifts substantially to lower 
temperatures. For BKT transition Kosterlitz predicted \cite{k74} that above $T_{c}$ the susceptibility diverges 
as $\chi\sim\exp[(2-\eta)b(T/T_{c}-1)^{-\nu}]$, with $\eta$=0.25, $b\approx$1.5 and $\nu$=0.5, and is
infinite below $T_{c}$. Our results suggest that in thermodynamic limit $N\rightarrow\infty$ there would 
be no divergence in $\chi$ at any finite temperature for the classical XY-model on Sierpi\' nski pyramid.

\section{Summary}

From our Monte Carlo simulation results we conclude that there is no finite temperature phase 
transition for the classical XY-model ($O(2)$ symmetry) on three-dimensional Sierpi\' nski pyramid 
(fractal dimension $D$=2). Since the heat capacity per site does not depend on the system 
size there can be no long range order at any finite temperature. Because the low-temperature 
helicity modulus decreases with increasing system size for closed boundary condition, and is zero within 
the error bars for open boundary condition, it must vanish in thermodynamic limit at any finite 
temperature. This implies no continuous finite temperature phase transition associated with 
unbinding of vortex strings \cite{ksw86} in which case the helicity modulus vanishes at transition 
temperature $T_{c}$ as a power law $\gamma(T)\propto|T-T_{c}|^{v}$ \cite{lt89}. Moreover there is 
no finite temperature Berezinskii-Kosterlitz-Thouless transition associated with unbinding of 
vortices and characterized by discontinuity in $\gamma(T)$ at $T_{c}$. These conclusions are 
supported by our results for linear magnetic susceptibility. The lack of finite-temperature 
long range order and the vanishing of spin stiffness/helicity modulus are the consequence of finite 
order of ramification of Sierpi\' nski pyramid: as an arbitrarily large bounded cluster of sites can 
be disconnected by cutting only the finite number of bonds thermal fluctuations drive  
helicity modulus to zero and destroy long range order.

\subsection{Acknowledgements}
We thank Professor S.~K.~Bose for many useful discussions. This work
was supported by the Natural Sciences and Engineering Research Council
(NSERC) of Canada. The work of M.~P.~was also supported in part
through an NSERC Undergraduate Student Research Award (USRA). 
\newpage
\begin{figure}
\begin{center}
\includegraphics[angle=270,width=10cm]{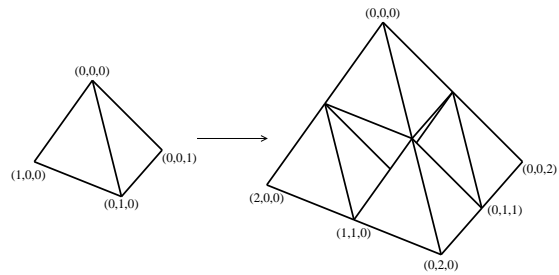}
\caption{\label{fig1}The first step in creating Sierpi\' nski pyramids.}
\label{fig1}
\end{center}
\end{figure}
\newpage
\begin{figure}
\begin{center}
\includegraphics[angle=0,width=10cm]{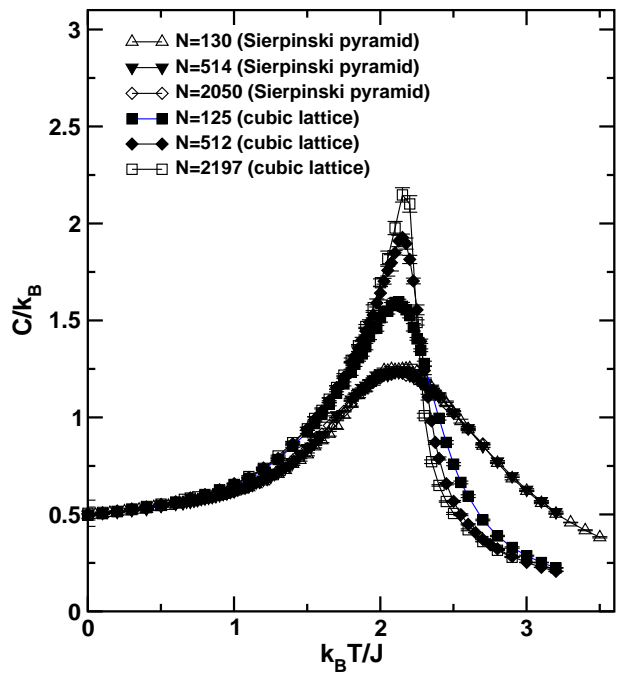}
\caption{\label{fig2}Calculated heat capacity per site as a function of temperature. The top three curves are 
obtained for the classical XY-model on cubic lattices while the results for Sierpi\' nski pyramids fall on the 
same curve within the error bars.} 
\label{fig2}
\end{center}
\end{figure}
\newpage
\begin{figure}
\begin{center}
\includegraphics[angle=0,width=10cm]{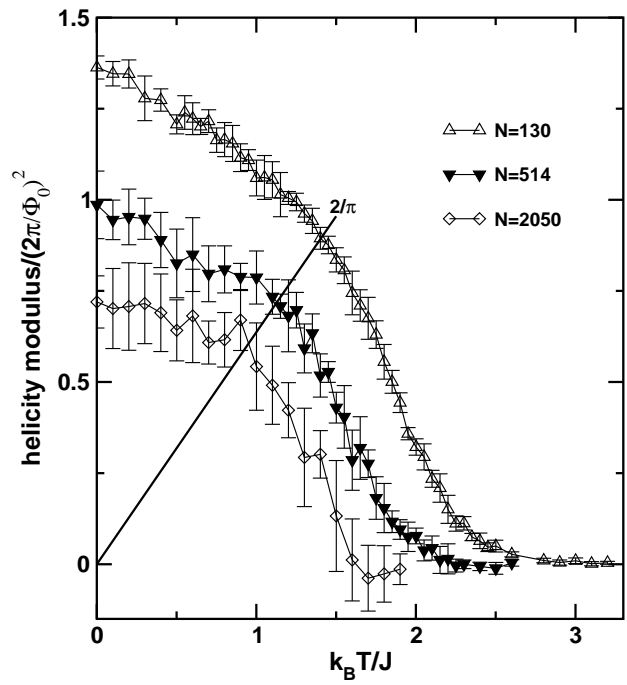}
\caption{\label{fig3}The helicity modulus $\gamma$ as a function of temperature for Sierpi\' nski 
pyramids of different sizes. The straight line gives the size of Kosterlitz-Thouless discontinuous 
jump in $\gamma$ at various temperatures.}
\label{fig3}
\end{center}
\end{figure}
\newpage
\begin{figure}
\begin{center}
\includegraphics[angle=0,width=10cm]{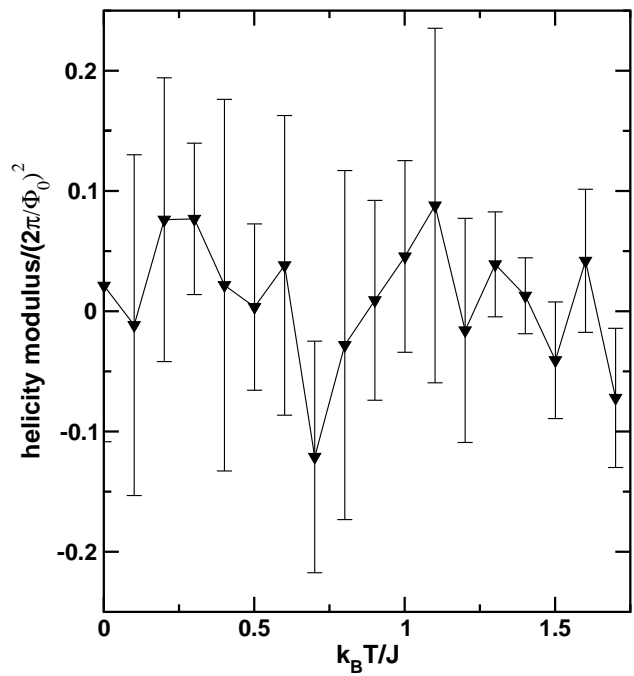}
\caption{\label{fig4}The helicity modulus $\gamma$ as a function of temperature for Sierpi\' nski 
pyramid of order $m$=5 (514 sites) obtained with the open boundary condition.}
\label{fig4}
\end{center}
\end{figure}
\newpage
\begin{figure}
\begin{center}
\includegraphics[angle=0,width=10cm]{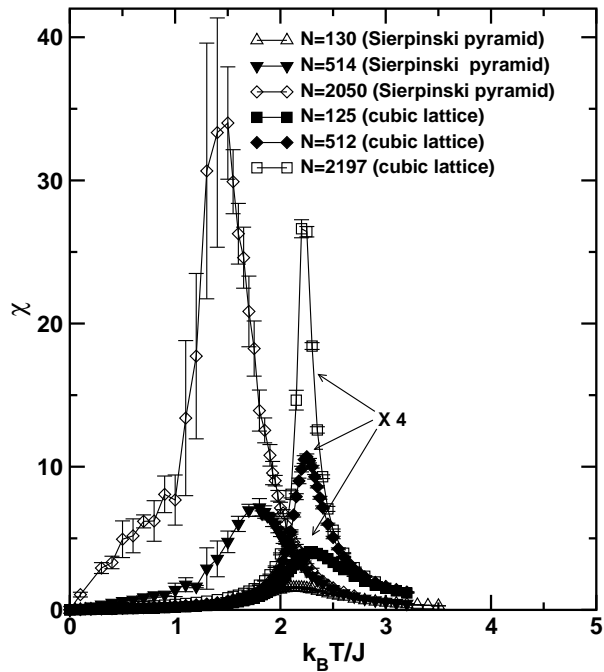}
\caption{\label{fig5}The linear susceptibility for for Sierpi\' nski pyramids and cubic lattices. Note that 
the values obtained for the cubic lattices have been increased by a factor of 4 in order to fit the scale 
set by the values obtained for the  pyramids.}
\label{fig5}
\end{center}
\end{figure}

\end{document}